\documentstyle[amssymb,aps,12pt]{revtex}
\begin{document}
\author{R. Radhakrishnan and M. Lakshmanan}
\author{Centre for Nonlinear Dynamics, Department of Physics,}
\author{Bharathidasan University, Tiruchirapalli - 620 024,}
\author{Tamil Nadu, India}
\title{Suppression and Enhancement of Soliton Switching During Interaction in
Periodically Twisted Birefringent Fiber}
\date{December 14, 1998 }
\maketitle

\begin{abstract}
Soliton interaction in periodically twisted birefringent optical fibers has
been analysed analytically with refernce to soliton switching. For this
purpose we construct the exact general two-soliton solution of the
associated coupled system and investigate its asymptotic behaviour. Using
the results of our analytical approach we point out that the interaction can
be used as a switch to suppress or to enhance soliton switching dynamics, if
one injects multi-soliton as an input pulse in the periodically twisted
birefringent fiber.

PACS Numbers 42.81Dp, 42.65Tg, 03.40kf
\end{abstract}

It is a well-known fact that single-core fiber supports two distinct modes
of propagation as a result of birefringence effect, which can be introduced
through twisting the fiber during the preform stage of formation or through
stress induced birefringence mechanism. For many years the propagation of
solitons in stress induced birefringent nonlinear Kerr media with reference
to optical fibers is an intensive research subject for theoretical as well
as experimental investigations [1]. The topic of propagation in twisted
birefringent optical fiber is also gaining considerable interest
theoretically and experimentally in recent times as most of the nonlinear
directional couplers are based on such fibers. One may mention the example
of the rocking rotator [1] which can be used as a switching device at
high-power [2] and as a filter at low-power [3].

Soliton propagation in the periodically twisted birefringent fiber is
usually described by using the coupled nonlinear Schr\"{o}dinger (CNLS)
family of equations [1,2]. However such CNLS equations are in general not
completely integrable. Interestingly if we assume the value of ellipticity
angle to be 35$^{{{}^{\circ }}}$ then the dynamics of soliton interaction in
a periodically twisted birefringent fiber coupler can be described by
coupled wave equations of the form [4]

\begin{eqnarray}
iq_{1z}+q_{1tt}+\rho q_1+\varkappa q_2+2\mu ( &\mid &q_1\mid ^2+\mid q_2\mid
^2)q_1=0,  \nonumber \\
iq_{2z}+q_{2tt}-\rho q_2+\varkappa q_1+2\mu ( &\mid &q_1\mid ^2+\mid q_2\mid
^2)q_2=0,  \label{e1}
\end{eqnarray}
where{\it \ q}$_{\text{1 }}${\it (z,t) }and{\it \ q}$_{\text{2 }}${\it (z,t)}
are slowly varying envelopes of two orthogonally polarized modes, z and t
are respectively the normalized distance and time and $\varkappa $ and $\rho 
$ are the normalized linear coupling constants caused by the periodic twist
of the birefringence axes and the phase-velocity mismatch from resonance
respectively. If the linear coupling constants are absent (that is, $%
\varkappa =\rho =0$), then one can easily recognize the system(1) to be the
celebrated integrable Manakov model [5]. The resulting Manakov equation is
receving renewed attention recently as it describes the effects of averaged
random birefringence on an orthogonally polarized pulse in a real fiber [6].%
{\bf \ }When the birefringene axes of fiber are periodically twisted during
the drawing process there is a periodic intensity exchange between the
orthogonally polarized modes [1-4] and it can be modelled by Eq.(1) if the
value of ellipticity angle is 35$^{\circ }$ [4]. The linear coupling length
where the maximum power is transfered from one mode to the other is $\pi /(2%
\sqrt{\varkappa ^2+\rho ^2})$. At the resonance wavelength, the linear
parameter $\rho =0$ \ and the linear coupling length increases to $\pi
/(2\varkappa )$. The schematic of the experimental apparatus used to observe
switching characters in the periodically twisted birefringent fiber is given
in Ref. [2]. Particularly the dependence of switching characters on the
input power, operating wavelength, twist magnitude and twist period are
described for example in Refs. [2] and [4].

Considerable attention has been paid in the literature [1,5,7,8] to study
soliton collision in the birefringent fiber. Particularly by using the
system (1) in the absence of linear coupling terms Manakov [5] pointed out
that during soliton collision their velocities and amplitudes (intensities)
do not change but the associated unit polarization vectors do change
provided they are neither parallel nor orthogonal (see also the paper of
Silmon-Clyde and Elgin [8] for a discussion in terms of Stokes vectors).
Further Menyuk [7] has observed at the value of the ellipticity angle 35$%
^{\circ }$, where the Manakov equation holds good, that a soliton of one
polarization when interacting with a switching pulse of the other
polarization does not develop a shadow and also does not change shape.
However very recently we have proved [9] by constructing the most general
two-soliton solution of the Manakov model that it has the property that
soliton in birefringent fiber can in general change its shape after
interaction due to change in intensity distribution among the modes
enenthough the total energy is conserved. In this letter we investigate the
implication of this property of the solitons when the additional effects due
to the periodic rotation of birefringence axes are included by constructing
the exact two-soliton solution of the system (1). In particular we point out
that interaction can be used as a switch to suppress or to induce soliton
switching, if we inject multisoliton as an input pulse in the periodically
twisted birefringent fiber.

The coupled system (1) reduces to the celebrated integrable Manakov model
[5],

\begin{eqnarray}
iq_{1Mz}+q_{1Mtt}+2\mu ( &\mid &q_{1M}\mid ^2+\mid q_{2M}\mid ^2)q_{1M}=0, 
\nonumber \\
iq_{2Mz}+q_{2Mtt}+2\mu ( &\mid &q_{1M}\mid ^2+\mid q_{2M}\mid ^2)q_{2M}=0,
\label{e2}
\end{eqnarray}
(the subscript M refers to the Manakov model) under the transformation [4],

\begin{eqnarray}
q_1 &=&\cos (\theta /2)e^{i\Gamma z}q_{1M}-\sin (\theta /2)e^{-i\Gamma
z}q_{2M},  \nonumber \\
q_2 &=&\sin (\theta /2)e^{i\Gamma z}q_{1M}+\cos (\theta /2)e^{-i\Gamma
z}q_{2M},  \label{e3}
\end{eqnarray}
where $\Gamma =(\rho ^2+\varkappa ^2)^{1/2}$ and $\theta =\tan
^{-1}(\varkappa /\rho )$. B\'{e}langer and Par\'{e} [10] and also briefly in
[1] have shown that the system (1) without linear self-coupling ($\rho =0)$
has simple solitary wave solutions exhibiting energy exchange between the
modes. By using the one-soliton solution of the Manakov model (2) in (3),
Potasek [4] has pointed out the possibility of periodic intensity exchange
between the orthogonally polarised modes {\it q}$_{\text{1 }}$and {\it q}$_{%
\text{2 }}$in the coupled system (1), when both the linear coupling
constants $\varkappa $ and $\rho $ are present. An interesting question
arises here when one considers multi-soliton solutions for the Manakov model
that admits both elastic and inelastic (shape changing) type of collisions
depending upon the initial conditions or arbitrary parameters as shown in
Ref. [9]. Then how does the switching and energy sharing properties get
modified for the multi-soliton solutions of the system (1)? We show in this
letter that indeed novel features in the intensity sharing and different
switching properties do arise when the most general two-soliton solution is
considered for (1).

For our analysis we make use of the general two-soliton solution of the
Manakov system (2), reported in [9], and obtain the corresponding
two-soliton solution of (1) through (3). It has the form :

\begin{eqnarray}
q_1 &=&\{{[\cos (\theta /2)e^{i\Gamma z}\alpha _1-\sin (\theta
/2)e^{-i\Gamma z}\beta _1]e^{\eta _1}+[\cos (\theta /2)e^{i\Gamma z}\alpha
_2-\sin (\theta /2)}  \nonumber \\
&&{e^{-i\Gamma z}\beta _2]e^{\eta _2}+[\cos (\theta /2)e^{i\Gamma z+\delta
_1}-\sin (\theta /2)e^{-i\Gamma z+\delta _1^{^{\prime }}}]e^{\eta _1+\eta
_1^{*}+\eta _2}+}  \nonumber \\
&&{[\cos }{(\theta /2)e^{i\Gamma z+\delta _2}-\sin (\theta /2)e^{-i\Gamma
z+\delta _2^{^{\prime }}}]e^{\eta _1+\eta _2+\eta _2^{*}}\}}/D_n,  \nonumber
\\
q_2 &=&\text{as q}_{\text{1 }}\text{above with the replacements }\cos
(\theta /2)\rightarrow \sin (\theta /2)\text{ and }  \nonumber \\
&&\sin (\theta /2)\rightarrow -\cos (\theta /2),
\end{eqnarray}
(the symbol * denotes complex conjugation). Here $D_n=1+\exp (\eta _1+\eta
_1^{*}+R_1)+\exp (\eta _1+\eta _2^{*}+\delta _0)+\exp (\eta _1^{*}+\eta
_2+\delta _0^{*})+\exp (\eta _2+$ $\eta _2^{*}+R_2)+\exp (\eta _1+\eta
_1^{*}+\eta _2+\eta _2^{*}+R_3)$ and $\eta _j=k_j(t+ik_jz)$, j=1,2. The
parameters $\exp (\delta _1)=(k_1-k_2)(\alpha _1\kappa _{21}-\alpha _2\kappa
_{11})/(k_1+k_1^{*})$ $(k_1^{*}+k_2),\exp (\delta _2)=(k_2-k_1)(\alpha
_2\kappa _{12}-\alpha _1\kappa _{22})/(k_2+k_2^{*})(k_1+k_2^{*}),\exp
(\delta _1^{^{\prime }})=(k_1-k_2)$ $(\beta _1\kappa _{21}-\beta _2\kappa
_{11})/(k_1+k_1^{*})(k_1^{*}+k_2),\exp (\delta _2^{^{\prime
}})=(k_2-k_1)(\beta _2\kappa _{12}-\beta _1\kappa _{22})/(k_2+k_2^{*})(k_1+$ 
$k_2^{*}),$ $\exp (\delta _0)=\kappa _{12}/(k_1+k_2^{*}),\exp (R_1)=\kappa
_{11}/(k_1+k_1^{*}),\exp (R_2)=\kappa _{22}/(k_2+$ $k_2^{*}),\exp (R_3)=\mid
k_1-k_2\mid ^2(\kappa _{11}\kappa _{22}-\kappa _{12}\kappa
_{21})/(k_1+k_1^{*})(k_2+k_2^{*})\mid k_1+k_2^{*}\mid ^2$ and $\kappa
_{ij}=\mu (\alpha _i\alpha _j^{*}+\beta _i\beta
_j^{*})(k_i+k_j^{*})^{-1},i,j=1,2.$ The six arbitrary complex parameters $%
\alpha _{1,}$ $\alpha _2,\beta _1,\beta _2,k_1$and $k_2$ determine the
amplitude, velocity and phase of the asymptotic soliton forms of (4). Now in
order to bring out the nature of the solitons of (1) and their interactions
including exchange of energy we carryout an asymptotic analysis of the
solution (4). To be specific we choose the arbitrary complex parameters $%
k_i,i=1,2$, as $k_{1I}>k_{2_I}$, $k_{1R}>0$ and $k_{2R}>0$ (here subscripts
I and R refer to the imaginary and real parts).

\underline{I. Limit z$\rightarrow -\infty $ : }As z$\rightarrow -\infty $ we
can identify two independent solitons denoted by soliton 1 and soliton 2
with the above choices of {\it k}$_1$and{\it \ k}$_2$. Soliton 1 will be
centered around $\eta _{1R}=k_{1R}(t-2k_{1I}z)\simeq 0$ (when $\eta
_{2R}\rightarrow -\infty $) and soliton 2 will be centered around $\eta
_{2R}=k_{2R}(t-2k_{2I}z)\simeq 0$ (when $\eta _{1R}\rightarrow \infty $).%
\newline
\underline{(a) Soliton 1 ($\eta _{1R}\simeq 0,\eta _{2R}\rightarrow -\infty
):$}

\begin{eqnarray}
&&q_1\cong [\cos (\theta /2)e^{i\Gamma z}A_{1M}^{1-}-\sin (\theta
/2)e^{-i\Gamma z}A_{2M}^{1-}]q^{1-},  \nonumber  \label{e5} \\
&&q_2\cong [\sin (\theta /2)e^{i\Gamma z}A_{1M}^{1-}+\cos (\theta
/2)e^{-i\Gamma z}A_{2M}^{1-}]q^{1-},  \label{e5ab}
\end{eqnarray}
where {\it q}$^{1-}=k_{1R}\exp (i\eta _{1I})\sec h(\eta _{1R}+R_1/2),\eta
_{1I}=k_{1I}t+(k_{1R}^2-k_{1I}^2)z,(A_{1M,}^{1-}$ $A_{2M}^{1-})=[\mu ($ $%
\alpha _1\alpha _1^{*}+\beta _1\beta _1^{*})]^{-1/2}(\alpha _{1,}\beta _1)$
and $\mid A_{1M}^{1-}\mid ^2+\mid A_{2M}^{1-}\mid ^2=1/\mu $ {\bf . }Here $%
(A_{1M,}^{1-}$ $A_{2M}^{1-})$ refers polarization unit vector of Manakov
one-soliton solution, which can be obtained from (5) by substituting $\theta
=\Gamma =0$, superscripts 1- denote soliton 1 at the limit z$\rightarrow
-\infty $ and subscripts 1 and 2 refer to the modes {\it q}$_1$and{\it q}$_2$%
. Eq.(5) exhibits the same form of one-soliton solution of (1) reported by
Potasek in the Ref. [4]. If we parametrize [8] the unit polarization vector
as $(A_{1M,}^{1-}$\ $A_{2M}^{1-})=(\cos (\theta _p^{1-})\exp (i\alpha
_{p1}^{1-}),\sin (\theta _p^{1-})\exp (i\alpha _{p2}^{1-}))$\ then we can
identify $\theta _p^{1-}$\ as the polarization angle and the phases $\alpha
_{p1}^{1-}$\ $\neq $\ $\alpha _{p2}^{1-}$\ corresponds to the state of
elliptical polarization. \newline
\underline{(b) Soliton 2 ($\eta _{2R}\simeq 0,\eta _{1R}\rightarrow \infty
): $}

\begin{eqnarray}
q_1 &\cong &[\cos (\theta /2)e^{i\Gamma z}A_{1M}^{2-}-\sin (\theta
/2)e^{-i\Gamma z}A_{2M}^{2-}]q^{2-},  \nonumber \\
q_2 &\cong &[\sin (\theta /2)e^{i\Gamma z}A_{1M}^{2-}+\cos (\theta
/2)e^{-i\Gamma z}A_{2M}^{2-}]q^{2-},  \label{e6ab}
\end{eqnarray}
where {\it q}$^{2-}=k_{2R}\exp (i\eta _{2I})\sec h(\eta
_{2R}+(R_3-R_1)/2),\eta _{2I}=k_{2I}t+(k_{2R}^2-k_{2I}^2)z,(A_{1M,}^{2-}$ $%
A_{2M}^{2-})=($a$_1/$a$_1^{*})$ c[$\mu (\alpha _2\alpha _2^{*}+\beta _2\beta
_2^{*})]^{-1/2}[(\alpha _{1,}\beta _1)\kappa _{11}^{-1}-(\alpha _{2,}\beta
_2)\kappa _{21}^{-1}],\mid A_{1M}^{2-}\mid ^2+\mid A_{2M}^{2-}\mid ^2$ = 1/$%
\mu $, in which a$_1=(k_1+k_2^{*})[(k_{1-}k_2)(\alpha _1^{*}\alpha _2+\beta
_1^{*}\beta _2)]^{1/2}$ and c=[ 1/$\mid \kappa _{12}\mid ^2-$ 1$/\kappa
_{11}\kappa _{22}$]$^{1/2}.$ It is interesting to note from (5) and (6) that
the form of $A_{1M}^{2-}$ and $A_{2M}^{2-}$ in (6) differs from the values $%
A_{1M}^{1-}$ and $A_{2M}^{1-}$ in (5) and the former contains more number of
parameters, eventhough (6) is an exact one-soliton solution of the system
(1) just like the solution (5). Further in the special case $\alpha
_1:\alpha _2=\beta _1:\beta _2,$ the form of (6) reduces to the form of (5)
with parameters specifying soliton 2. So Eq.(6) may be considered as the
most general one-soliton solution form of (1).

\underline{II. Limit z$\rightarrow \infty $ : } We now analyse the form of
the solitons after interactions as z$\rightarrow \infty :$\newline
\underline{(a) Soliton 1 ($\eta _{1R}\simeq 0,\eta _{2R}\rightarrow \infty
): $}

\begin{eqnarray}
q_1 &\cong &[\cos (\theta /2)e^{i\Gamma z}A_{1M}^{1+}-\sin (\theta
/2)e^{-i\Gamma z}A_{2M}^{1+}]q^{1+},  \nonumber \\
q_2 &\cong &[\sin (\theta /2)e^{i\Gamma z}A_{1M}^{1+}+\cos (\theta
/2)e^{-i\Gamma z}A_{2M}^{1+}]q^{1+},  \label{e7ab}
\end{eqnarray}
where {\it q}$^{1+}=k_{1_R}\exp (i\eta _{1I})\sec h(\eta
_{1R}+(R_3-R_2)/2),(A_{1M,}^{1+}A_{2M}^{1+})=($a$_2/$a$_2^{*})$ c[$\mu
(\alpha _1\alpha _1^{*}+\beta _1\beta _1^{*})]^{-1/2}[(\alpha _{1,}\beta
_1)\kappa _{12}^{-1}-(\alpha _{2,}\beta _2)\kappa _{22}^{-1}],\mid
A_{1M}^{1+}\mid ^2+\mid A_{2M}^{1+}\mid ^2$ = 1/$\mu $, in which a$%
_2=(k_2+k_1^{*})[(k_{1-}k_2)(\alpha _1\alpha _2^{*}+\beta _1\beta
_2^{*})]^{1/2}.$ Note that A$_{1M}^{1+}$ $\neq $A$_{1M}^{1-}$ and A$%
_{2M}^{1+}\neq A_{2M}^{1-}$, except when $\alpha _1:\alpha _2=\beta _1:\beta
_2$, corresponding to pure elastic collision in the Manakov model [9]. 
\newline
\underline{(b) Soliton 2 ($\eta _{2R}\simeq 0,\eta _{1R}\rightarrow -\infty
):$}

\begin{eqnarray}
q_1 &\cong &[\cos (\theta /2)e^{i\Gamma z}A_{1M}^{2+}-\sin (\theta
/2)e^{-i\Gamma z}A_{2M}^{2+}]q^{2+},  \nonumber \\
q_2 &\cong &[\sin (\theta /2)e^{i\Gamma z}A_{1M}^{2+}+\cos (\theta
/2)e^{-i\Gamma z}A_{2M}^{2+}]q^{2+},  \label{e8ab}
\end{eqnarray}
where {\it q}$^{2+}=k_{2R}\exp (i\eta _{2I})\sec h(\eta
_{2R}+R_2/2),(A_{1M}^{2+}$ $A_{2M}^{2+})=[\mu (\alpha _2\alpha _2^{*}+\beta
_2\beta _2^{*})]^{-1/2}(\alpha _{2,}\beta _2)$ and $\mid A_{1M}^{2+}\mid
^2+\mid A_{2M}^{2+}\mid ^2=1/\mu .$ Here also A$_{1M}^{2+}$ $\neq $A$%
_{1M}^{2-}$ and A$_{2M}^{2+}\neq A_{2M}^{2-}$, unless $\alpha _1:\alpha
_2=\beta _1:\beta _2$.

Now we recognise from (5-8) that not only the phase-factors but also the
overall shapes of the solitons get modified (due to the intensity
redistribution among the solitons) after undergoing interaction when the two
coupled one-solitons (5) and (6) move from z$\rightarrow -\infty $ to z$%
\rightarrow \infty $ as shown in Eq.(7) and Eq.(8). To facilitate the
understanding of the above behaviour with reference to the optical soliton
switching between the orthogonally polarised modes, it is convenient to
obtain the oscillating parts of the intensities associated with the
asymptotic forms (5-8) as

\begin{eqnarray}
%TCIMACRO{\QOVERD| | {q_l(z,t)}{q^{n\mp }(z,t)} }
%BeginExpansion
{q_l(z,t) \overwithdelims|| q^{n\mp }(z,t)}%
%EndExpansion
^2 &=&\mid A_{lM}^{n\mp }\mid ^2\cos ^2(\theta /2)\text{+}\mid A_{jM}^{n\mp
}\mid ^2\sin ^2(\theta /2)\text{ + (-1)}^l\mid A_{lM}^{n\mp }\mid \mid
A_{jM}^{n\mp }\mid  \nonumber \\
&&\sin (\theta )\cos (2\Gamma z+\phi ^{n\mp }),\text{ l,j=1,2 (l}\neq \text{%
j), z}\rightarrow \mp \infty ,  \label{e9}
\end{eqnarray}
where $\phi ^{n\mp }=\tan ^{-1}%
%TCIMACRO{\QOVERD( ) {A_{1MI}^{n\mp }}{A_{1MR}^{n\mp }}}
%BeginExpansion
{A_{1MI}^{n\mp } \overwithdelims() A_{1MR}^{n\mp }}%
%EndExpansion
-\tan ^{-1}%
%TCIMACRO{\QOVERD( ) {A_{2MI}^{n\mp }}{A_{2MR}^{n\mp }}}
%BeginExpansion
{A_{2MI}^{n\mp } \overwithdelims() A_{2MR}^{n\mp }}%
%EndExpansion
.$ The presence or absence of the last term involving the factor $\cos
(2\Gamma z+\phi ^{n\mp })$ plays a crucial role in the switching behaviour
of solitons as demonstrated below.

As we have mentioned before the values of A$_{jM}^{n-}$ (j,n=1,2) change to
new values A$_{jM}^{n+}$ due to the collision between two co-propagating
solitons, namely soliton 1 and soliton 2, without violating the condition $%
\mid A_{1M}^{n\mp }\mid ^2+\mid A_{2M}^{n\mp }\mid ^2=1/\mu $. The amount of
change (A$_{jM}^{n+}-$A$_{jM}^{n-}$) can be estimated by assigning suitable
values to the arbitrary parameters k$_1$,k$_2$,$\alpha _1$,$\alpha _2$,$%
\beta _1,$ and $\beta _2,$ appearing in the expressions for A$_{jM}^{n\mp }$
(j,n=1,2). Further from (9) it is obvious that depending upon the values of $%
\mid $A$_{jM}^{n\mp }\mid ,$ the nature of switching dynamics supported by
the system (1) also changes. The change in $A_{jM}^{n\mp }$\ can be
parmeterized as $(\cos (\theta _p^{n\mp })\exp (i\alpha _{p1}^{n\mp }),\sin
(\theta _p^{n\mp })\exp (i\alpha _{p2}^{n\mp }))$, where as long as the
phases $\alpha _{p1}^{n\mp }\neq \alpha _{p2}^{n\mp }$\ the state of
polarization is preserved during interaction. Therefore in general without
affecting the state of polarization, the switching dynamics can be changed
just by changing $A_{jM}^{n\mp }$with the help of the polarization angle. In
the following we briefly discuss the different changes which can occur in
the intensity exchange between {\it q}$_1$ and {\it q}$_2$ modes with
respect to soliton 1 and soliton 2 due to the above mentioned collision by
considering $%
%TCIMACRO{\QOVERD| | {q_l}{q^{n-}}}
%BeginExpansion
{q_l \overwithdelims|| q^{n-}}%
%EndExpansion
^2$ and $%
%TCIMACRO{\QOVERD| | {q_l}{q^{n+}}}
%BeginExpansion
{q_l \overwithdelims|| q^{n+}}%
%EndExpansion
^2$ defined in (9) for each l=1,2 and n=1,2 values.

\underline{Case 1 :} All the $\mid A_{jM}^{n\mp }\mid $'s (j,n=1,2) are
nonzero. In this case due to the presence of $\cos (2\Gamma z+\phi ^{n\mp })$
term on the right hand side of (9), there is a periodic intensity switching
which is always present in both the solitons and in both the components
before as well as after the interaction. Of course the conservation
relations $\mid A_{1M}^{n-}\mid ^2+\mid A_{2M}^{n-}\mid ^2=\mid
A_{1M}^{n+}\mid ^2+\mid A_{2M}^{n+}\mid ^2=1/\mu $ for the total intensity
are always valid. However the switching dynamics appearing before and after
interaction is not similar in form, due to the condition A$_{jM}^{n+}$ $\neq 
$A$_{jM}^{n-}$, j,n=1,2 , except when $\alpha _1:\alpha _2=\beta _1:\beta _2$
as mentioned before, giving rise to a partial suppression or enhancement of
the periodically varying intensities.

\underline{Case 2 :} Any one of the $\mid A_{jM}^{n\mp }\mid $'s is zero and
others are nonzero. For example if $\mid A_{1M}^{1+}\mid \sim 0,$
corresponding to the condition $\alpha _2(k_{2+}k_2^{*})(\alpha _1\alpha
_2^{*}+\beta _1\beta _2^{*})=\alpha _1(\mid \alpha _2\mid ^2+\mid \beta
_2\mid ^2)(k_1+k_2^{*}),$ then the switching in the intensity of the soliton
1 gets fully suppressed in both the modes {\it q}$_1$ and {\it q}$_2$, while
it persists for the other soliton. This is illustrated in Fig. 1a for the
choosen parameters values, namely k$_1=1+i,k_2=2-i,\alpha _1=\beta _1=\beta
_2=1,\alpha _2=(39+i80)/89(\simeq \exp (i64^{{{}^{\circ }}})),\rho =0.25$
and $\varkappa =0.5,$ for which $\mid A_{1M}^{1-}\mid \sim 0.7$,$\mid
A_{1M}^{2-}\mid \sim 0.5$,$\mid A_{2M}^{1-}\mid \sim 0.7$,$\mid
A_{2M}^{2-}\mid \sim 0.86$,$\mid A_{1M}^{1+}\mid \sim 0.06$,$\mid
A_{1M}^{2+}\mid \sim 0.7$, $\mid A_{2M}^{1+}\mid \sim 0.99$ and $\mid
A_{2M}^{2+}\mid \sim 0.7$, satisfying the condtion that $\mid
A_{1M}^{1+}\mid \sim 0.$ Similar phenomenon can be seen if $\mid
A_{2M}^{1+}\mid \sim 0$ instead of $\mid A_{1M}^{1+}\mid \sim 0.$ But if we
choose $\mid A_{jM}^{2+}\mid \sim 0$ (j=1 or j=2) the periodic intensity
exchange with respect to soliton 2 will be suppressed, while it persists in
soliton 1. On the other hand, if $\mid A_{jM}^{n-}\mid \sim 0$ (j=1 or j=2;
n=1 or n=2), then there is no switching in the intensity of soliton n before
interaction, but the switching appears after interaction in that soliton and
so there is an inducement of switching due to the inteaction. Thus the
interaction itself acts as a switch to suppress or to enhance the switching
dynamics.

\underline{Case 3 :} Any two of the $\mid A_{jM}^{n\mp }\mid $'s are zero
and others are nonzero without violating the conservation conditions. For
concreteness, let $\mid A_{1M}^{n+}\mid $ $\sim 0$ $(n=1,2),$ which implies
the condition $\alpha _1\sim \alpha _2\sim 0.$ It implies that $\mid
A_{1M}^{n-}\mid $ (n=1,2) should also simultaneously vanish. Consequently
there is no switching between the modes {\it q}$_1$ and {\it q}$_2$ either
before or after interaction and there will be only inelastic (shape
changing) scattering as discussed for the Manakov model in Ref. [9]. Similar
observations can also be made if one makes any two of the $\mid A_{jM}^{n\mp
}\mid $'s to vanish. However, one can identify the interesting possibility
of switching existing in soliton 1 only before interaction, which gets
interchanged with soliton 2 after interaction, by allowing one of the two $%
\mid A_{jM}^{2-}\mid $'s and another one of $\mid A_{jM}^{1+}\mid $'s
simultaneously to take the value zero, corresponding to the condition $[\mid
k_1+k_2^{*}\mid ^2(\mid \alpha _1\mid ^2+\mid \beta _1\mid ^2)(\mid \alpha
_2\mid ^2+\mid \beta _2\mid ^2)]=[(k_1+k_1^{*})(k_2+k_2^{*})$ $\mid \alpha
_1\alpha _2^{*}+\beta _1\beta _2^{*}\mid ^2].$ Figure 1b, for the choosen
parameters namely $k_1=1+i0.1,k_2=1-i0.1,\alpha _1=0.86+i0.5,\alpha
_2=0.5+i0.86,\beta _1=0.7+i0.72,$ $\beta _2=0.44+i0.9,$ and $\rho =\varkappa
=0.25$, for which $\mid A_{1M}^{1-}\mid \sim 0.7$,$\mid A_{1M}^{2-}\mid \sim
0.05$,$\mid A_{2M}^{1-}\mid \sim 0.7$,$\mid A_{2M}^{2-}\mid \sim 0.99$,$\mid
A_{1M}^{1+}\mid \sim 0.04$,$\mid A_{1M}^{2+}\mid \sim 0.7$, $\mid
A_{2M}^{1+}\mid \sim 0.99$ and $\mid A_{2M}^{2+}\mid \sim 0.7$ satisfying
the above condtion, shows that the interaction induces the periodic
intensity exchange between the two modes of soliton 2 while it suppresses
the switching dynamics in soliton 1.

To conclude, by studying the interaction between two co-propagating solitons
in the periodically twisted birefringent fiber with reference to soliton
switching, we have observed several possible ways to use interaction as a
switch to suppress or to induce the switching dynamics. The basic underlying
mechanism of such possibilities is the inelastic (shape changing) nature of
the soliton interaction which arises essentially due to changes in the
polarization angle and so in the overall amplitude of the solitons. Since
Mollenauer et al [8] have demonstrated polarization scattering by
soliton-soliton collision, it shoud also\ possible to experimentally study
the phenomena described in this paper by using specially fabricated optical
fibers. These possibilities should have important ramifications in nonlinear
switching devices like the rocking rotator.

{\it Acknowledgments : }The authors are grateful to Dr. Jarmo Hietarinta for
useful discussions. This work has been supported by the Deparment of Science
and Technology, Government of India.

\bigskip

\part{Figure Caption}

Fig.1 : Typical evolution of the intensity profiles $\mid q_1\mid ^2$ and $%
\mid q_2\mid ^2$ of the two-soliton solution (4) (a) showing the suppression
in switching between the two modes of s1 soliton and (b) showing the
suppression in switching of s1 soliton and enhancement in s2 soliton (while
undergoing a large phase shift) for the parameter values given in the text.

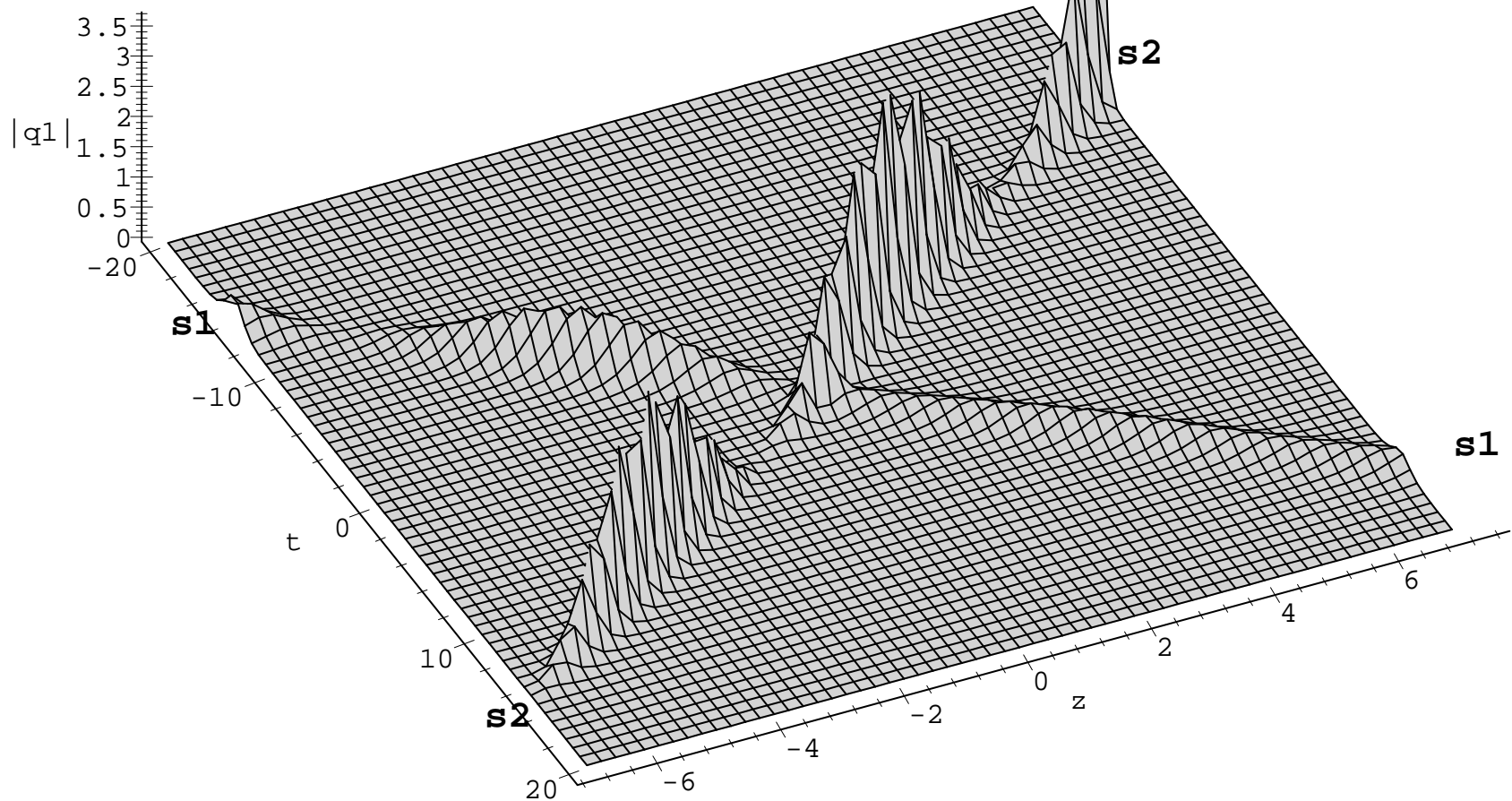
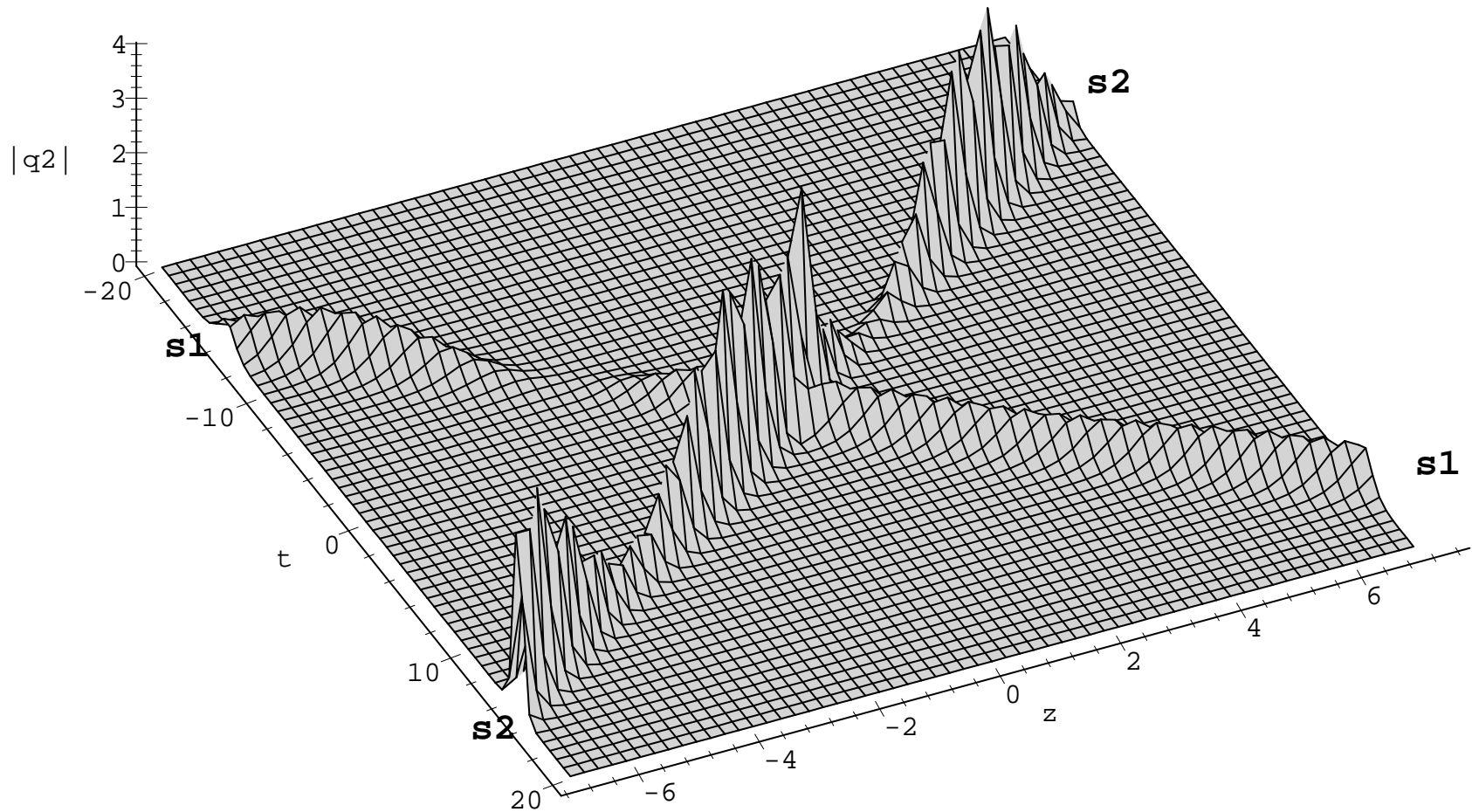
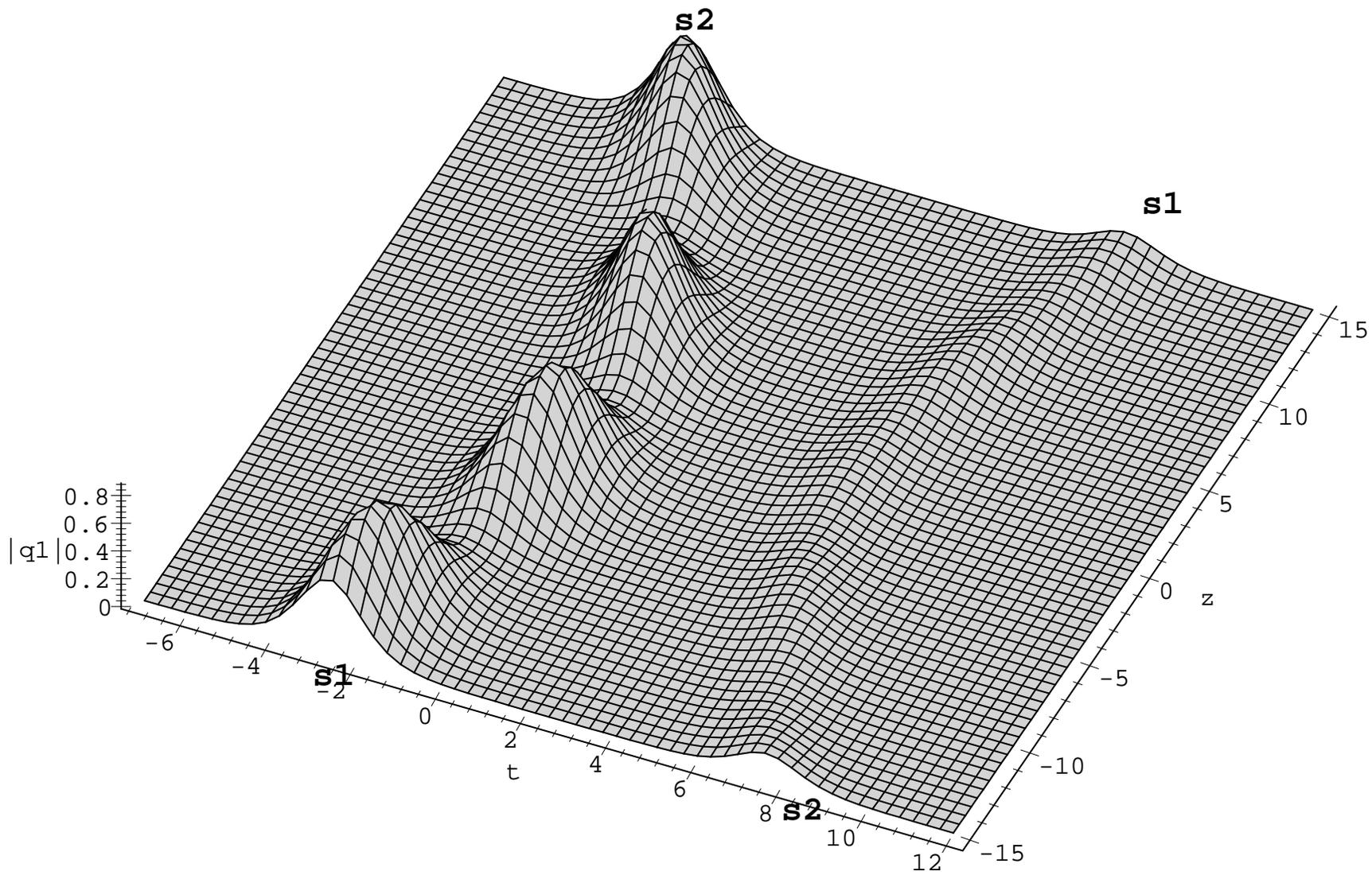
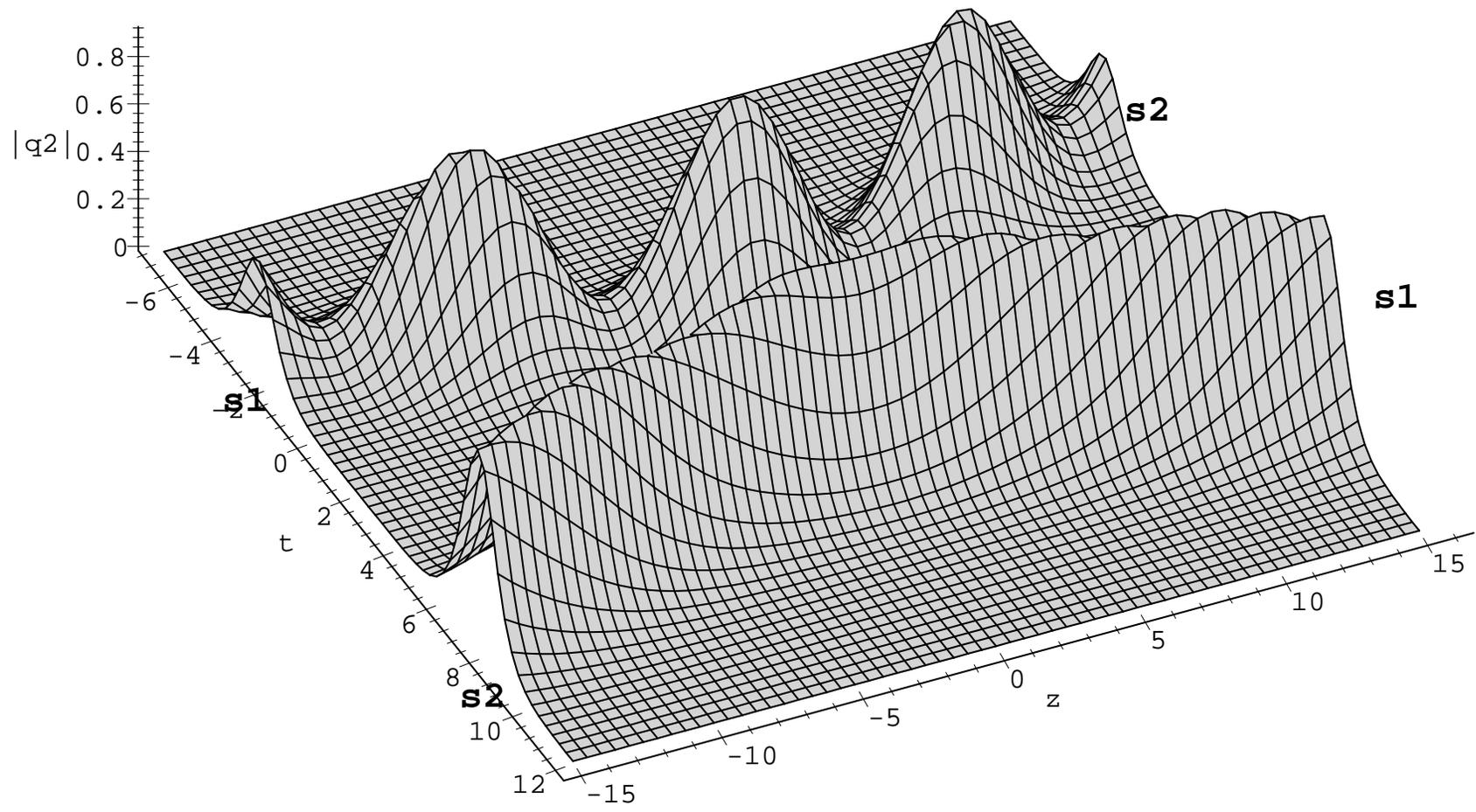
\end{document}